# Liquid money or hard cash?
## Drowning into granular material.

Franco Bagnoli,
Dept. of Physics and Astronomy and Center for the Study of Complex Dynamics, University of Florence, Italy
Via G. Sansone, 1 50019 Sesto Fiorentino (FI) Italy
franco.bagnoli@unifi.it

In British English, the term "hard cash" refers to the form of payment using coins or bill, rather than cheques or credit or money transfer. In American English, it is often prefixed by the adjective "cold". On the contrary, in Italian the equivalent expression "denaro liquido" can be literary translated as "liquid money". In French the expression is equivalent with the additional factor, with respect to the rest of this discussion, that money becomes "argent".

We have therefore two very different points of view: Is money hard and cold, or rather liquid and "jingling" ("moneta sonante")? As usual, we shall investigate this topic starting from some comics about the duck family. We all know that Uncle Scrooge loves to dive into his money "like a porpoise" [1], so for him, money is indeed liquid.

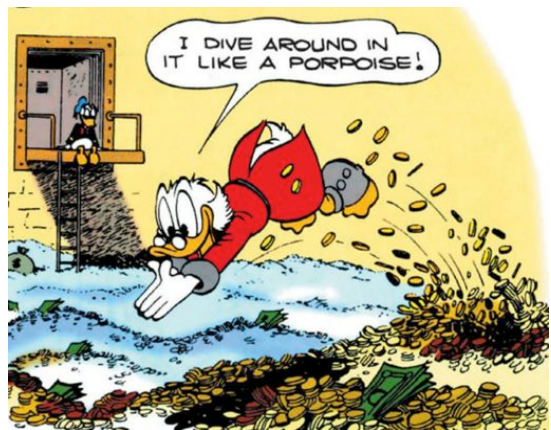
Figure 1. Swimming into a sea of coins [1]

However, near the end of the same story, the Beagle Boys succeeded in stealing his money, but they are lured in enjoying it "like the tightwad used to do".

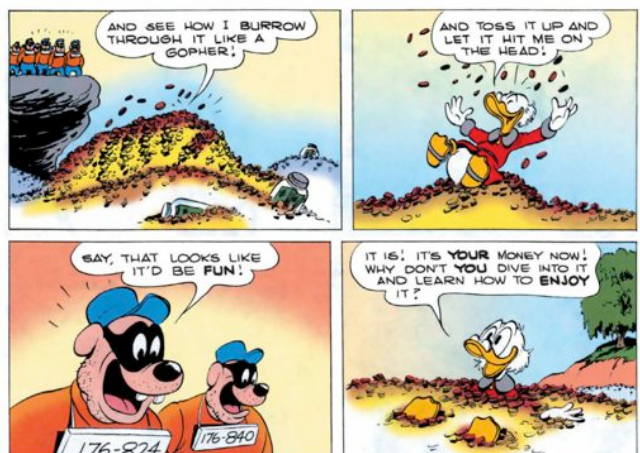
Figure 2. Luring the Beagle Boys [1].

The burglars try to plunge into the mass of coins but, as expected, they crash against a solid body.

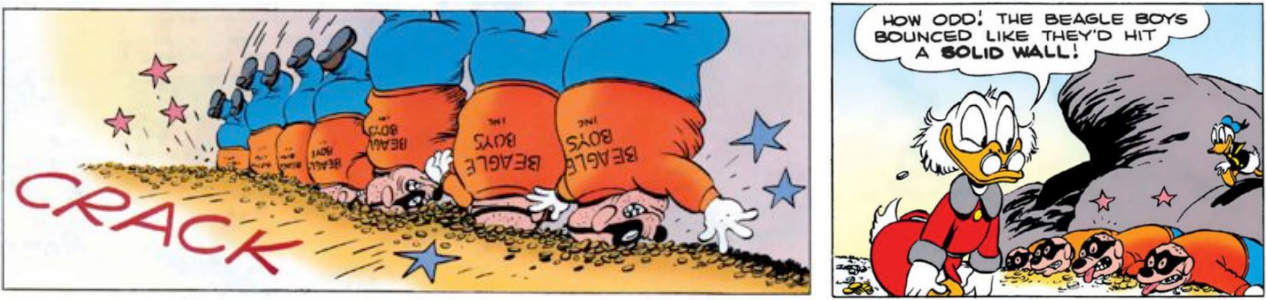

Figure 3. Cold, hard money [1]

Indeed, a granular material can behave both as a solid and as a liquid, depending on the density of its kinetic energy. We all know that transferring a granular material like rice, sugar or coffee from a container to another is not a trivial task. These materials behave quite differently from liquids: if the container is inclined, the mass of grains behaves at first like a rigid body and stays compact. Passing a critical angle, it suddenly (partially) liquefies and originates an avalanche. However, even without an academic knowledge, we all know what to do to pour a granular material from one container into another: fluidify it by continuously shaking.

The cohesion of a granular material is due to the same kind of forces that originates friction, i.e., a stick-and-slip dynamics. It can be visualized using two combs.

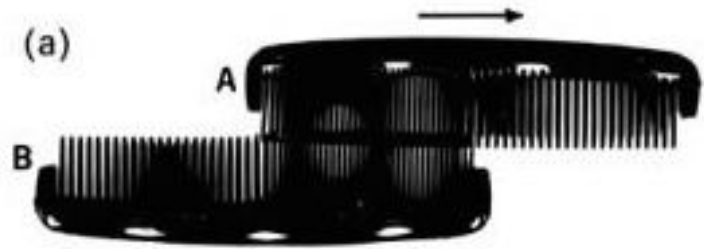

Figure 4. Stick-and-slip dynamics of combs [2].

The usual distinction from static and dynamic friction law is sufficient to qualitatively explain the behavior of a granular material.

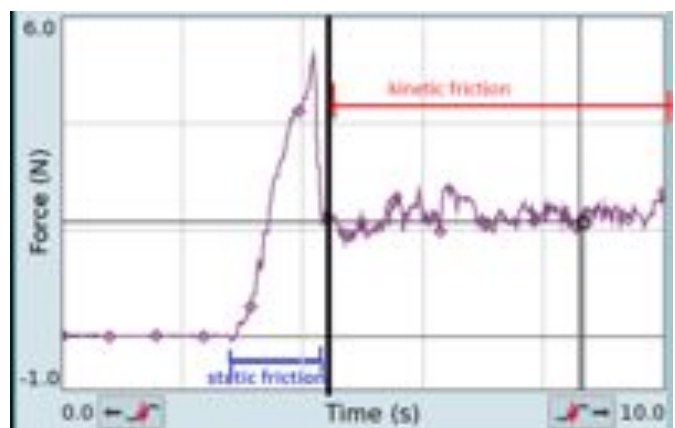

Figure 5. Static and kinetic friction experiment [3].

When the static friction is overcome, the grain start moving, and the dynamic friction is unable to bring the system at rest, when the external force is kept constant. So, if one provides continuous energy, the granular material behaves like a sort of viscous fluid, although the viscosity is not proportional to the velocity, and has a nonlinear transition to infinite viscosity for low enough velocity. The system dissipates energy, so we have a different scenario with respect to a molecular fluid. We can speak of two different temperatures: the microscopic and the macroscopic one, the latter roughly identified by the macroscopic kinetic energy of grains, even though the velocity distribution of a granular material is in general not Gaussian, due to the continuous conversion of the macroscopic energy into the microscopic one. The shape of the velocity distribution is not universal: it depends on how energy is injected (for instance, by vibrating the fluid with a given amplitude and velocity shape) and how it is dissipated (by collision, by viscous drag from air, and so on) [4].

Another aspect of granular materials is the arching phenomena. The distribution of stresses inside a granular system is not isotropic. Due to the nonlinear behavior of the friction coefficient, the material self-organizes so that all stresses above threshold are removed, and what is left are percolating chains of forces.

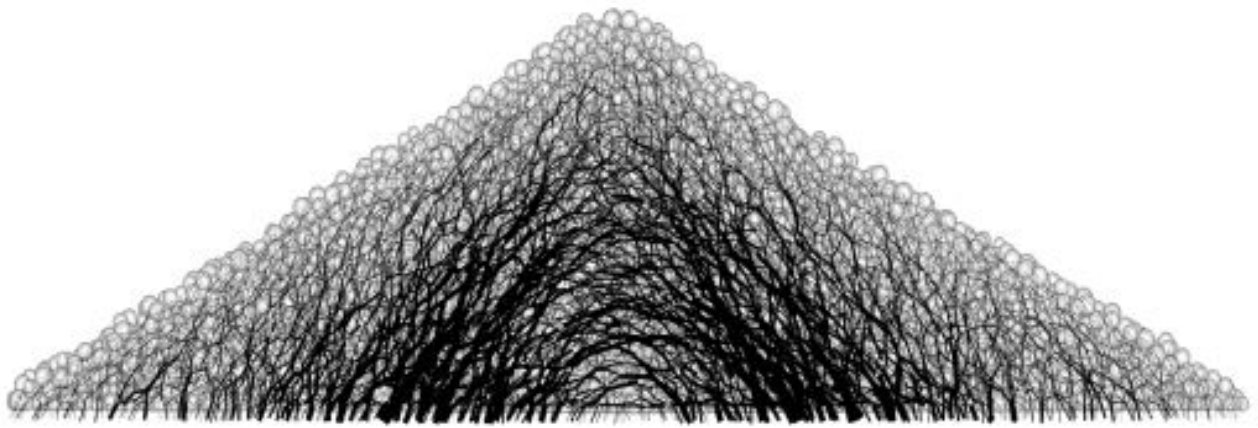

Figure 6. Force chains in a discrete model of a sand heap – arching effect [5].

This is the origin of the stability of sand piles, and of the fact that granular materials do not obey to the Stevino's law: the pressure at the bottom of a silo is not what can be expected: the pressure stops increasing once the grain is higher than roughly the silo diameter [6]. For higher levels, the load is redirected to the walls, and this can make the silos to break. Indeed, when loading silos, care is taken to let the granular material keep the liquid state long enough, using appropriate geometries.

The arching phenomenon is not always deleterious: hourglasses behaves differently from water clocks and produce a stationary flow of material due to the fact that the arching makes the pressure at bottom independent of the sand load.

However, arching can also clog granular material in silos.

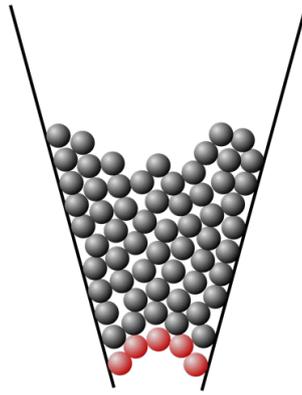

Figure 7. Jamming during discharge of granular material due to arch formation [7].

This inconvenient is also experienced by the Beagle Boys in "Cash Flow" by Don Rosa [8], although in their interpretation the clogging is only attributed to bills.

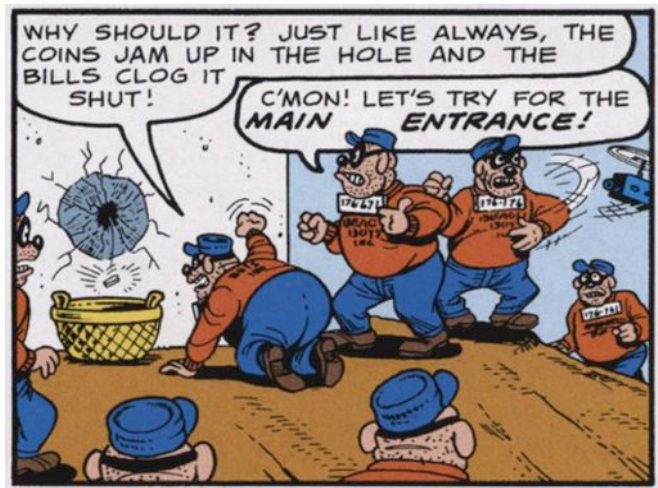

Figure 8. Hard cash clogging [8].

In this comic, a scientist, trying to remove fumes from cabbage, invented two rays: one that removes friction, and another that cancels the inertial mass, but not the gravitational one (Einstein would disagree on the feasibility of this latter ray).

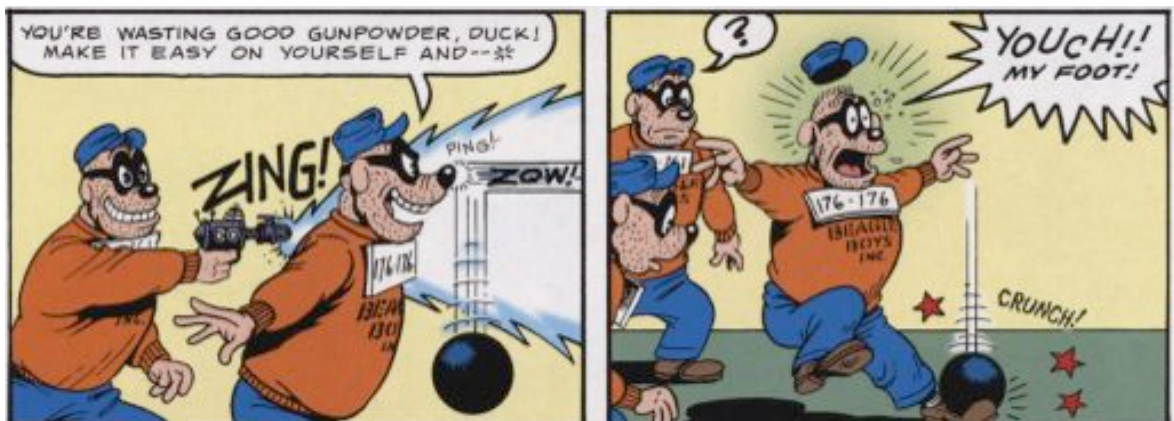

Figure 9. Removing inertial but not gravitational mass [8]. The actual trajectory of the cannon ball would be quite different: without inertial mass it would immediately fall on the floor.

They succeed in entering Scrooge's Money Bin, but the old duck manages in fluidizing the coins by removing their friction, so that now the granular material behaves like much as a superfluid (otherwise the burglars could have used buckets to carry the liquid money).

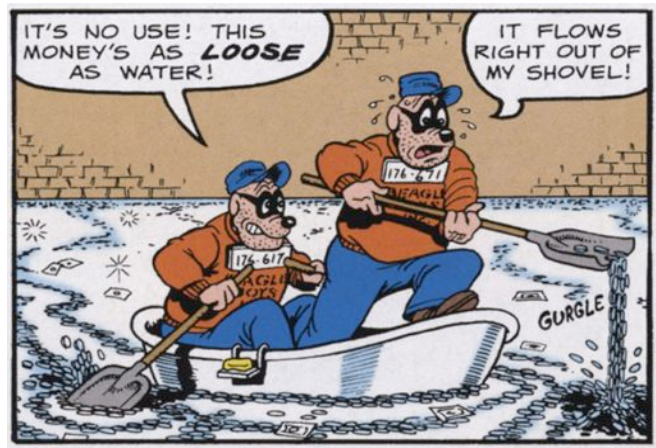
Figure 10. Liquid money [8].

The Beagle Boys could finally enjoy swimming among dollars, but for some strange reason, they appear to drown in them.

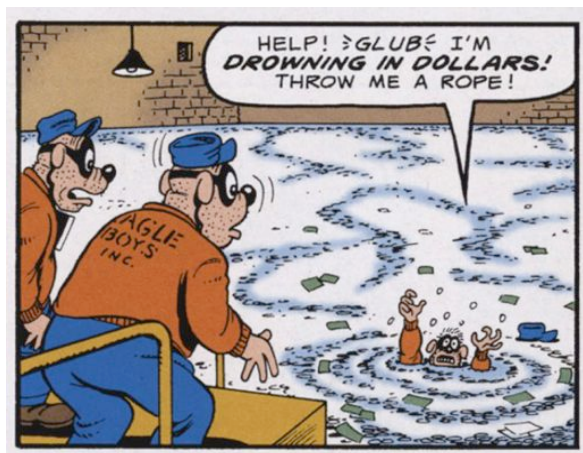
Figure 11. Drowning by dollars [8].

This is quite strange, since the money retains its mass (the second ray was not used, and in any case it only cancels inertial mass), and the density of iron is about seven times that of water. Even considering that the random packing lowers the density by a factor of about one half (due to interstitial spaces), the Beagle Boy should float like a piece of iron in a bowl of mercury.

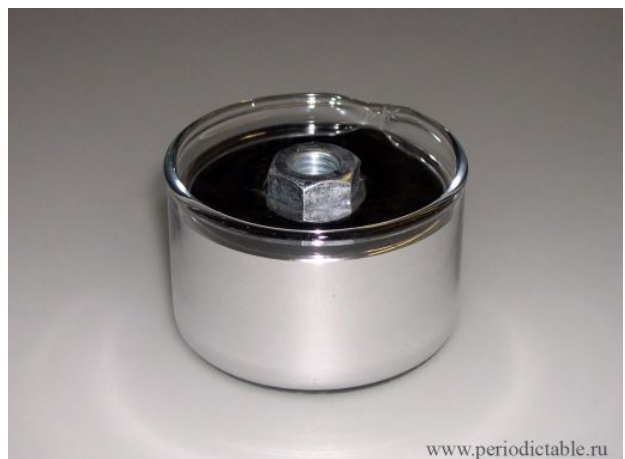
Figure 12. Iron floating on mercury [9].